\documentstyle[12pt]{article}

\input epsf

\begin{document}
\begin{titlepage}
\today          \hfill
\begin{center}
\hfill    OITS-692 \\

\vskip .05in

{\large \bf
Charged Higgs decays 
in models with singlet neutrino in large
extra dimensions}
\footnote{This work is supported by DOE Grant DE-FG03-96ER40969.}
\vskip .15in
K. Agashe \footnote{email: agashe@oregon.uoregon.edu},
N.G. Deshpande \footnote{email: desh@oregon.uoregon.edu},
G.-H. Wu \footnote{email: wu@dirac.uoregon.edu}

\vskip .1in
{\em
Institute of Theoretical Science \\
5203 University
of Oregon \\
Eugene OR 97403-5203}
\end{center}

\vskip .05in

\begin{abstract}
In models with large extra dimensions and TeV scale quantum gravity,
small (Dirac) neutrino masses can be obtained naturally if the right-handed 
neutrino propagates in the extra dimensions. In this scenario,
with $2$ Higgs doublets,
we show that
the decay of charged Higgs into {\em left-handed} charged lepton (say $\tau$)
and right-handed neutrino can be significantly
enhanced, with $O(1)$ branching ratio,
due to the large number of
Kaluza-Klein states of the right-handed neutrino.
Since $\tau$'s from 
the standard decay of charged Higgs are {\em right-handed},
the above novel
charged Higgs decay can provide a 
distinctive signature of these models
at hadron/lepton colliders. 
Similarly, top quark decays to $b \mu^+ \nu$,
$b \tau^+ \nu$ through virtual charged Higgs can be enhanced.

\end{abstract}

\end{titlepage}

\newpage
\renewcommand{\thepage}{\arabic{page}}
\setcounter{page}{1}


The framework of large extra dimensions and TeV scale quantum gravity
\cite{add} has been 
extensively studied over the last two years.
The idea is that there are $\delta$ additional spatial dimensions of size
$R$ in which gravity lives whereas the SM particles are confined to
the usual $4$ dimensions ($4D$). The effective $4D$ Planck scale, $M_{Pl}
\sim 2.4 \times 10^{18}$ GeV, is related to the $(4 + \delta) D$ 
``fundamental''
Planck scale, $M_{\ast}$, by
\begin{equation}
M_{Pl}^2 \sim R^{\delta} M_{\ast} ^{\delta+2}.
\label{funda}
\end{equation}
Thus, if the extra dimensions are large $(R \gg
M^{-1}_{\ast})$, then it is possible that
$M_{\ast} \sim$ TeV so that the ultraviolet (UV) cut-off for the 
quantum
corrections to the Higgs boson (mass)$^2$ is also $\sim$ TeV.
Therefore, the gauge hierarchy problem is ``solved''. The size of the
extra dimensions can be as large as $\sim$ mm for $\delta = 2$.

References \cite{aetal, ddg} showed how small neutrino masses can
naturally be obtained in this framework even though there is no fundamental
mass scale much larger than the weak scale to implement the usual
seesaw mechanism for neutrino masses. 
Since the right-handed (RH)
neutrino is a singlet under the SM gauge group, it can live
in the extra dimensions
\cite{fp, ds, dk, bcs, others}. 
In this paper, we study charged Higgs decays to
$\tau \bar{\nu}$ in this scenario.
For simplicity, consider the case of a single
extra dimension labeled by $y$ ($x$ labels the usual $4D$). 
A massless Dirac fermion $N$
which is a singlet lives in $5D$. The $\Gamma$ 
matrices in $5D$ can be written as
\begin{equation}
\Gamma ^{\mu} = 
\left(
\begin{array}{cc}
0 & \sigma ^{\mu} \\
\bar{ \sigma } ^{\mu} & 0
\end{array}
\right), 
\Gamma ^5 = 
\left(
\begin{array}{cc}
i & 0 \\
0 & -i
\end{array}
\right).
\end{equation}
The Dirac spinor $N$ in the Weyl basis can be written as
\begin{equation}
N = 
\left(
\begin{array}{c}
\psi \\
\bar{\chi}
\end{array}
\right),
\end{equation}
where $\psi$ and $\chi$ are $2$-component complex spinors
(with mass dimension $2$).
The $5D$ kinetic term for $N$ is
\begin{equation}
S_{\hbox{free}} = \int d^4 x dy \; \bar{N} \left( \Gamma ^{\mu}
\partial _{\mu} + \Gamma ^5 \partial _y \right) N.
\label{kinetic}
\end{equation}
In the effective $4D$ theory, $N$ appears as a tower of Kaluza-Klein 
(KK) states:
\begin{equation}
\psi = \sum _n \frac{1}{\sqrt{R}} \psi ^{(n)} (x)
e^{iny / R},
\label{kk}
\end{equation}
where $\psi ^{(n)}$ are $4D$ states.   
Similarly, $\chi$ has a KK tower, $\chi ^{(n)}$.
Thus, in $4D$, Eq. (\ref{kinetic}) becomes
\begin{equation}
S_{\hbox{free}} = \int d^4x \sum _n \left[ \bar{\psi}
^{(n)} \bar{\sigma}^{\mu}
\psi
^{(n)} + \bar{\chi}
^{(n)} \bar{\sigma}^{\mu}
\chi
^{(n)} + \left( \frac{n}{R} \psi ^{(n)} \chi ^{(n)} + \hbox{h.c.} 
\right) \right],
\label{kinetic4D}
\end{equation}
where $n / R$ is the Dirac mass for the KK states.
If we assign $N$ the opposite lepton number from the usual lepton doublet
$l = ( \nu , e )$ (which of course lives on a ``$3$-brane''
localized at say $y = 0$), then the interaction
between $l$ and $N$ which conserves lepton number is   
\begin{equation}
S_{\hbox{int}} = \int d^4x \; \frac{\lambda}
{\sqrt{M_{\ast }}} \; l (x) h^{\ast} (x)
\psi (x, y = 0),
\label{Lint}
\end{equation}
where $h$ is the SM Higgs doublet (with hypercharge
$-1/2$)
and $\lambda$ is dimensionless.

In the effective $4D$ theory, Eq. (\ref{Lint}) becomes
\begin{equation}
S_{\hbox{int}} = \int d^4x \; \sum _n \; \frac{\lambda}{\sqrt{R M_{\ast}}}
l (x) h^{\ast} (x) \psi ^{(n)} (x).
\end{equation}
Using Eq. (\ref{funda})
we get 
\begin{equation}
S_{\hbox{int}} = \int d^4x \; \sum _n \; \lambda
\frac{M_{\ast}}{M_{Pl}} l (x) h^{\ast} (x) \psi ^{(n)} (x).
\label{Sint4D}
\end{equation}
This can be easily generalized to the case of
$\delta$ extra dimensions resulting in the same effective $4D$
coupling (for now,
we assume that the singlet neutrino lives in the same $\delta$
extra dimensions as the graviton).
Thus, we see that the neutrino Yukawa coupling is
suppressed by volume of extra dimensions 
so that the Dirac mass for the SM neutrino is
\begin{equation}
m = \frac{\lambda}{\sqrt{2}}
\frac{M_{\ast}}{M_{Pl}} v,
\label{m}
\end{equation}
where $v \approx 246$ GeV is the Higgs vev.
 
From Eqns. (\ref{kinetic4D}) and (\ref{Sint4D}),
we see that $\chi ^{(0)}$ decouples and is exactly massless. 
The full mass matrix for $\nu$, $\psi ^{(0)}$, $\psi ^{(n)}$ and $\chi ^{(n)}$ 
$(n = ..,-2, -1,1,2,..)$ in the case of one extra dimension is
\begin{equation}
{\cal L}_{\hbox{mass}} = \nu _{+} M \nu _{-}^T
\end{equation}
with
\begin{equation}
M = 
\left( 
\begin{array}{ccccc}
m & m & m & m &... \\
0 & 1/R & 0 & 0 &... \\
0 & 0 & -1/R & 0 & ... \\
0 & 0 & 0 & 2/R & ... \\
... & ... & ... & ... & ... 
\end{array}
\right),
\label{M}
\end{equation}
where
\begin{equation}
\nu _{+} = \left( \nu, \chi^{(1)}, \chi^{(-1)},
\chi^{(2)}, ... \right)
\end{equation}
and
\begin{equation}
\nu _{-} = \left( \psi ^{(0)}, \psi ^{(1)}, \psi ^{(-1)},
\psi ^{(2)}, ... \right).
\end{equation}
In the limit $m \ll 1/R$, we have to a good approximation a
Dirac fermion $( \nu, \psi ^{(0)} )$
with mass $m$ (see,
however, Eq. (\ref{mnu1})) 
and Dirac fermions $(\psi ^{(n)}, \chi ^{(n)})$ $(n = ..,-2, -1,1,2,..
)$, with
masses $n / R$ with the mixing between
$\nu$ and $\chi ^{(n)}$ given by $\sim m R / n \ll 1$.

Since we will study charged Higgs decays to $\tau \bar{\nu}$,
we are interested in the case $m^2 \sim \Delta m^2 _{\hbox{\small atm}}
\sim 10^{-3}- 10^{-2}$ (eV)$^2$ (or larger, see later) as
indicated by atmospheric
neutrino 
oscillations \cite{superk}. 
For $\delta = 2$,
even for $M_{\ast} \sim 10$ TeV, $1/R$ is quite small $\sim
0.01 - 0.1$ eV so that with $m^2 \sim \Delta m^2 _{\hbox{\small atm}}$,
we get $m R \sim 1$, i.e., the
above approximation is no longer valid since there is $O(1)$ mixing between
the SM neutrino and the light KK neutrinos. Also, due to this large
mixing, this scenario might
be ruled out by the success of standard 
theory of Big Bang Nucleosynthesis (BBN) since too many singlet neutrinos
will be thermal during nucleosynthesis in the early universe
\cite{aetal, ds, bcs}.
Hence, in this paper, we will mostly
consider $\delta \geq 3$ for
which $1/R \gg 0.1$ eV.
Also, in the supernova $1987$a core, neutrinos (of all flavors)
undergo coherent oscillations to  
KK neutrinos leading to energy loss at an unacceptable rate
if $m^2 \stackrel{>}{\sim} 10^{-3}$
(eV)$^2$ and $1/R \stackrel{<}{\sim} 10$ keV \cite{bcs}. 
We will show that the effect of RH neutrino in extra dimensions
on charged Higgs decays can be significant even in
the parameter space for which this constraint is 
satisfied.

To generalize the above to the case of three SM neutrinos,
we can add three singlets $N_i$ $(i = 1,2,3)$.

\vskip .1in

{\large \bf Charged Higgs decays}

\vskip .1in

From Eq. (\ref{Sint4D}), it is clear that the neutral Higgs has a coupling
$\lambda \; M_{\ast} / M_{Pl}$ to each of the RH neutrino
KK states, $\psi ^{(n)}$,
so that the neutral Higgs decay to neutrinos can be enhanced by the large
number of KK states \cite{aetal, mw}. 

Similarly,
in a $2$-Higgs-doublet model, 
the charged Higgs decay to (say) $\tau_L$ and
(RH) neutrino can be enhanced. 
For simplicity,
we assume (as, for example, in a supersymmetric extension of the SM)
that the Higgs doublet with hypercharge $ - 1/2$ (denoted by
$H_1$) couples only to RH
up-type quarks and neutrinos
(i.e., gives mass to up-type quarks and neutrinos)
whereas the hypercharge $ + 1/2$ doublet (denoted by $H_2$)
couples only to
RH charged leptons and down-type quarks -- we will refer to this model
({\em in} $4D$) as $2$-Higgs-Doublet Model II ($2$HDM-II, for short).
The ratio of the vev's of $H_1$ and $H_2$ is denoted by
$\tan \beta$ as usual. 
Neglecting the $\tau$ mass in the phase space integral
and also neglecting terms suppressed by $\sim m_{\tau}^2 / m_H^2$
in the matrix element, 
the decay width to LH $\tau$ is given by
\begin{eqnarray}
\Gamma \left( H ^- \rightarrow \tau _L \psi \right) & \approx &
\frac{m_H}{8 \pi} \left( \frac{m}{v} \right)^2 \cot  ^2 \beta  \;
(m_H R )^{\delta} \; x _{\delta} \nonumber \\
 & \approx & \frac{m_H}{8 \pi} \left( \frac{m}{v} \right)^2
\cot ^2 \beta
\left(
\frac{m_H}{M_{\ast}} \right) ^{\delta} \left( \frac{M_{Pl}}{M_{\ast}} 
\right)^2 \; x_{\delta} ,
\label{tauL}
\end{eqnarray}
where the factor $(m_H R )^{\delta}$ counts (roughly)
the number of RH neutrino
KK states
lighter than the charged Higgs
and we have used Eq. (\ref{funda}) in the second line. 
We have replaced the sum over KK states by an integral, $\sum _n
\rightarrow S_{\delta -1} n^{\delta -1} dn$ (where $S_{\delta -1} =
2 \pi^{\delta/2} / \Gamma( \delta /2 )$ is the surface area of a unit-radius 
sphere in $\delta$ dimensions). \footnote{For $\delta$ extra dimensions,
$n$ is really $\sqrt{ \sum _{i=1} ^{\delta} n_i^2 }$, where
$n_i$ is the momentum (in units of $\sim 1/ R$) in the $i^{\hbox{\small th}}$
extra dimension.}
This together with the phase space integral results in the factor
$x _{\delta}$ 
which is 
given by
\begin{equation}
x_{\delta} \approx \frac{2 \pi ^{\delta/2}}{\Gamma (\delta/2)} 
\left( \frac{1}{\delta} 
- \frac{2}{\delta + 2} + \frac{1}{\delta + 4} \right).
\label{x}
\end{equation}

As usual the charged Higgs also decays to RH $\tau$ through the $\tau$
Yukawa coupling:
\begin{eqnarray}
\Gamma \left( H^- \rightarrow \tau _R \bar{\nu} \right) & \approx &
\left[
\frac{m_H}{8 \pi} \left( \frac{m_{\tau}}{v} \right)^2 \tan ^2 \beta
\right] \nonumber \\
 & & \times
\frac{1}{N^2} \; \left[ 1 + \frac{m^2}{M_{\ast}^2} \frac{
m_H^{\delta -2}}{M_{\ast}^{\delta -2}} 
\frac{M_{Pl}^2}{M_{\ast}^2} 
\frac{2 \pi ^{\delta/2}}{\Gamma (\delta/2)}
\left( \frac{1}{\delta -2}
- \frac{2}{\delta} + \frac{1}{\delta + 2} \right) \right].
\label{tauR}
\end{eqnarray}
This decay is also affected by the presence of the RH neutrino in
extra dimensions as follows. The SM neutrino (weak eigenstate) is dominantly
the lightest neutrino with mass $\sim m$, but it has a small mixture
$(\sim m R /n)$ of the heavier neutrinos (see Eq. (\ref{M})). This 
mixing
introduces a ``normalization factor'' $N$ (in the second line
of the above equation) given by
\begin{equation}
N^2 \approx 
1 + \sum_n \left( \frac{mR}{n} \right) ^2 \approx
1 + \frac{m^2}{M_{\ast}^2} \frac{M_{Pl}^2}{M_{\ast}^2}
\frac{2 \pi ^{\delta/2}}{\Gamma (\delta/2)}  \frac{1}{\delta -2},
\label{norm1}
\end{equation}
where the sum over KK states is up to the UV cut-off, $M_{\ast}$.
Also, due to this mixing,
the charged Higgs also decays into RH $\tau$ and the 
{\em heavier} neutrinos which
have mass $\sim n/R$ and hence a different phase space
(compared to
the standard decay)  --
this accounts for the extra factor 
in
the numerator 
in the second line of Eq. (\ref{tauR}) (here the KK states are summed up to
the threshold of the decay). For $\delta = 2$, we have to replace
$1/ (\delta - 2)$ by $\ln \left( m_H M_{Pl} / M^2_{\ast} \right)$ and
$\ln \left( M_{Pl} / M_{\ast} \right)$ in the numerator
and the denominator, respectively. 

It is clear from Eqns. (\ref{tauR}) 
and (\ref{norm1}) that the decay width to RH $\tau$ is actually
{\em reduced} 
(as long as $m_H <
M_{\ast}$)
compared to that in $2$HDM-II:
the term in the first line
of Eq. (\ref{tauR}) is the decay width in $2$HDM-II.

Also, the mass
of the lightest neutrino is modified due to the normalization factor
\cite{aetal}:
\begin{equation}
m_{\nu} \approx \frac{m}{N}
\approx \frac{m} { \sqrt{
1 + \frac{ m^2 }{ M_{\ast}^2 } \frac{ M_{Pl}^2 }{ M_{\ast}^2 }
\frac{ 2 \pi ^{\delta/2} }{ \Gamma (\delta/2) }  \frac{1}{\delta -2} } }.
\label{mnu1}
\end{equation}
Thus, the physical neutrino mass cannot be increased arbitrarily
by increasing $m$ (or, in other words, by increasing $\lambda$):
the upper limit (for given $M_{\ast}$ and $\delta$) is
\begin{equation}
m_{\nu}^{\hbox{\scriptsize max}} \approx \frac{ M_{\ast}^2 }{ M_{Pl} }
\sqrt{ \frac{ \Gamma (\delta / 2) \; (\delta - 2) }
{ 2 \pi ^{ \delta / 2 } } },
\label{mnumax}
\end{equation}
where as before, for $\delta = 2$, $\delta -2$ is replaced by
$1 / \ln \left( M_{Pl} / M_{\ast} \right)$.
Using Eqs. (\ref{norm1}), (\ref{mnu1}) and (\ref{mnumax}), we get
the useful relations 
\begin{equation}
N^2 \approx 1 + \left( \frac{m}{m_{\nu}^{\hbox{\scriptsize max}}} \right)^2
\label{norm2}
\end{equation}
and
\begin{equation}
m_{\nu} \approx \frac{m} {\sqrt{ 1 + 
\left( \frac{m}{m_{\nu}^{\hbox{\scriptsize max}}} \right)^2 } }.
\label{mnu2}
\end{equation}

The ratio of decay widths to LH and RH $\tau$'s is
\begin{equation}
x_{LR} \equiv \frac{\Gamma \left( H^- \rightarrow \tau _L \psi \right)}
{\Gamma \left( H^- \rightarrow \tau _R \bar{\nu} \right)} \sim
\cot ^4 \beta \left( \frac{m}{m_{\tau}} \right) ^2 
\left(
\frac{m_H}{M_{\ast}} \right) ^{\delta} \left( \frac{M_{Pl}}{M_{\ast}} 
\right)^2,
\label{ratioLR}
\end{equation}
where, for simplicity, we have dropped the 
normalization and phase space factors.
For the parameter values $m^2 _{\nu} \sim 10^{-2}$ (eV)$^2$
(as applicable to solutions to the atmospheric neutrino anomaly),
$M_{\ast} \sim 2$ TeV, $\delta = 3$,
$\tan \beta \sim 2$ and $m_H \sim 200$ GeV,
we get $x_{LR} \sim 2 \times 10^5$. \footnote{For these values
of the parameters, the decay width to $\tau _R$ is suppressed
by a factor $\sim 10$ relative to $2$HDM-II.}
We see that the decay to LH $\tau$ can dominate
over the decay to RH $\tau$: 
the Yukawa 
coupling to $\tau _L$ and an {\em individual} KK neutrino is tiny, but there
is a large multiplicity factor. It should be possible to measure
the $\tau$ polarization at hadron/lepton colliders and thus these two
decays can be distinguished.

If the charged Higgs is lighter than the top quark
\footnote{The LEP2 limit on the charged Higgs mass is $\sim 80$
GeV \cite{aleph} and for
$\tan \beta \sim 100$, charged Higgs masses up to $\sim 120$ GeV
are excluded by CDF \cite {cdf}. 
}, then the only other
significant decay mode is $\bar{c} s$ which is at most comparable to
$\tau _R \bar{\nu}$ (and is unaffected by the 
presence of singlet neutrino in extra dimensions).
In this case, since the decay width to LH $\tau$ $\propto m_H ^{\delta}$,
$x_{LR}$ is a bit smaller than the
above estimate, but it can still be $\gg 1$.
Then, we see that the
BR for the decay mode $\tau _L \psi$ can be close to 1.
 
If the charged Higgs is heavier than the top quark, then the decay mode
$\bar{t} b$ (and maybe $W^- h^0$) is kinematically allowed.
The decay width $H^- \rightarrow \bar{t} b$ is
\begin{eqnarray}
\Gamma \left( H^- \rightarrow \bar{t} b \right) & \approx & 3 \;
\frac{m_H}{8 \pi} \left[ 
\left[ \left( \frac{m_t}{v} \right) ^2 \cot ^2 \beta
+ \left( \frac{m_b}{v} \right) ^2 \tan ^2 \beta 
\right] \left( 1 - x_t^2 - x_b^2 \right) - 
\frac{4 m_t^2 m_b^2}{v^2 m_H^2} \right] \nonumber \\
 & & \times \sqrt{ \left[ 1 - \left( x_t +
x_b \right) ^2 \right] \left[ 1 - \left(x_t - x_b \right)^2 \right] }, 
\label{tb}
\end{eqnarray}
where the factor of $3$ is for the number of colors
and $x_t \equiv m_t / m_H$, $x_b \equiv m_b / m_H$. 

The 
decay width $ H^- \rightarrow W^- h^0$ 
is given by
\begin{equation}
\Gamma \left( H^-
\rightarrow W^- h^0
\right) \approx \frac{1}{2 \pi} \cos ^2 (\beta - \alpha)
\frac{p_W^3}{v^2},
\label{hW}
\end{equation}
where $\alpha$ is the mixing angle of the
CP-even neutral Higgs scalars 
and $p_W$ is the magnitude of the $3-$momentum
of the $W$ boson in the rest frame of the charged Higgs:
\begin{equation}
p_W = \frac{1}{2 m_H} \sqrt{ \left[ m_H^2 - \left( m_W + m_h \right)^2 
\right]
\left[ m_H^2 - \left( m_W - m_h \right)^2 \right] }.
\end{equation}

Thus, for small $\tan \beta$ and dropping the phase space factors,
the ratio of decay widths $H^- \rightarrow \tau _L \psi$ and
$H^- \rightarrow \bar{t} b$ is
\begin{equation}
\frac{\Gamma \left( H^- \rightarrow \tau _L \psi \right)}
{\Gamma \left( H^- \rightarrow \bar{t} b \right)} \sim \frac{1}{3}
\left( \frac{m}{m_t} \right) ^2 \left(
\frac{m_H}{M_{\ast}} \right) ^{\delta} \left( \frac{M_{Pl}}{M_{\ast}} 
\right)^2
\label{ratiotauLtb}
\end{equation}
from which we see that even if $H^-$ is heavier than the top quark, 
the decay to $\tau _L \psi$ 
can dominate.
For the above values of parameters the ratio
is $O(100)$ so that the BR
for $H^- \rightarrow \tau _L \psi$ 
can be $O(1)$. 

{\bf Numerical results}

The charged Higgs signatures for
the scenario of singlet neutrino
in large extra dimensions depend on the parameters $M_{\ast}$,
$m^2$, $m_H$, $\tan \beta$ and $\delta$
as follows.

First of all, using Eq. (\ref{mnumax}), we see that
for $M_{\ast} \approx 20$ TeV, 
we get
$\left( m_{\nu}^{\hbox{\scriptsize max}} \right) ^2 \approx 3 \times 10^{-3}$
(eV)$^2$
(roughly independent of $\delta$). Therefore,
to get
$m_{\nu} ^2 \sim \Delta m _{\hbox{\small atm}} ^2 
\sim 10^{-3} - 10^{-2}$ (eV)$^2$,
we require
$M_{\ast} \stackrel{>}{\sim}
20$ TeV.
Conversely, if $M_{\ast} \sim$ (a few) TeV, then
$m_{\nu}^2 \ll \Delta m ^2 _{\hbox{\small atm}}$ (irrespective of
the value of $m$). So, we consider three cases: {\bf (i)} $M_{\ast}
\sim 20$ TeV, {\bf (ii)} $M_{\ast} \gg 10$ TeV and {\bf (iii)} $M_{\ast} \sim$
(a few) TeV.

{\bf Case (i)} $M_{\ast}
\sim 20$ TeV so that $m_{\nu}^{\hbox{\scriptsize max}}
\sim \sqrt{\Delta m^2 _{\hbox{\small atm}}}$

\begin{figure}
\centerline{\epsfxsize=1\textwidth \epsfbox{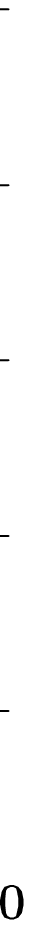}}
\caption{The contour plot for $BR(H^-\rightarrow \tau
\bar{\nu})$ (solid lines)
(left figure) and $\tau$ polarization asymmetry, $A _{\tau}$
(see Eq. (\ref{Adef})) (right figure)
in the $m_H$-$\tan \beta$ plane for $3$ extra dimensions,
$M_{\ast} =20$ TeV and $m \approx 0.1$ eV which corresponds
to $m_{\nu}^2 \approx 1.8
\times 10^{-3}$  (eV)$^2$ (case (i) (a) of text).
The prediction for $BR(H^-\rightarrow \tau \bar{\nu})$
in a
$2$HDM-II
is shown with
broken lines (left figure).
The decay modes considered for
the total charged Higgs decay width are $\tau \bar{\nu}$, $\bar{t} b$ and
$W^- h^0$.
We assume $m_{h^0} = 110$ GeV and the
mixing angle of the CP-even
neutral
Higgs scalars,
$\alpha = \pi /6$.
}
\protect\label{BRpolsmall}
\end{figure}

(a) $m \sim \sqrt{ \Delta m ^2 _{\hbox{\small atm}} }$
so that $N ^2 \sim O(1-10)$ (see Eq. (\ref{norm2}))
and $m_{\nu} \sim 
\sqrt{ \Delta m ^2 _{\hbox{\small atm}} }$ 
(see Eq. (\ref{mnu2})).
In Fig.\ref{BRpolsmall}, we plot the BR for $H^- \rightarrow \tau \bar{\nu}$
and the polarization of $\tau$ in the $\tan \beta - m_H$ plane
\footnote{This is the result of the full computation, Eqns.
(\ref{tauL}), (\ref{tauR}), (\ref{tb}) and (\ref{hW}).}
as a typical example of this
case: $\delta =3$, $m \approx 0.1$ eV and
$M_{\ast} = 20$ TeV. This gives $N^2 \approx 5.5$
and $m _{\nu} ^2 \approx 1.8 \times 10^{-3}$ (eV)$^2$. The
$\tau$ polarization asymmetry is defined as
\begin{equation}
A _{\tau} = \frac{\Gamma \left( H^- \rightarrow \tau _L \psi\right) -
\Gamma \left( H^- \rightarrow \tau _R \bar{\nu}\right) }
{\Gamma \left( H^- \rightarrow \tau _L \psi\right) +
\Gamma \left( H^- \rightarrow \tau _R 
\bar{\nu} \right) }.
\label{Adef}
\end{equation}

In this example, 
the enhancement of decay width to LH $\tau$ is not
so strong (especially for 
large $\tan \beta$, see Eqs. (\ref{ratioLR})
and (\ref{ratiotauLtb})) while the decay width to RH $\tau$ is suppressed 
(as compared to that in a $2$HDM-II) 
by a factor of $O(1-10)$ (see Eq. (\ref{tauR})).
Thus, the total BR to $\tau$
is smaller than 
that in $2$HDM-II.
However, for small $\tan \beta$, 
it is still possible that a significant fraction of 
$\tau$'s from Higgs decay are
LH.
These features can be seen in Fig.\ref{BRpolsmall}. 

\begin{figure}
\centerline{\epsfxsize=1\textwidth \epsfbox{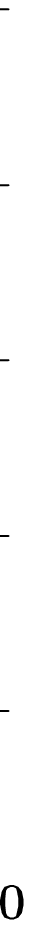}}
\caption{The contour plot for $BR(H^-\rightarrow \tau
\bar{\nu})$ (solid lines)
(left figure) and $\tau$ polarization asymmetry, $A _{\tau}$
(see Eq. (\ref{Adef})) (right figure)
in the $m_H$-$\tan \beta$ plane for $3$ extra dimensions,
$M_{\ast} =20$ TeV and $m \approx 3$ eV which corresponds
to $m_{\nu}^2 \approx 2.2 \times 10^{-3}$ (eV)$^2$ (case (i) (b) of text).
The prediction for $BR(H^-\rightarrow \tau \bar{\nu})$
in a
$2$HDM-II
is shown with
broken lines (left figure). The decay modes considered for
the total charged Higgs decay width are $\tau \bar{\nu}$, $\bar{t} b$ and
$W^- h^0$.
We assume $m_{h^0} = 110$ GeV and the
mixing angle of the CP-even
neutral
Higgs scalars,
$\alpha = \pi /6$.
}
\protect\label{BRpollarge}
\end{figure}

(b) $m ^2 \sim$ (few eV)$^2 
\gg \Delta m ^2 _{\hbox{\small atm}}$ so that
$N^2 \gg 1$ 
and $m_{\nu} \approx m_{\nu}^ {\hbox{\scriptsize max}}
\sim \sqrt{ \Delta m ^2 _{\hbox{\small atm}} }$. 
A typical example of this case is shown in 
Fig.\ref{BRpollarge}: $\delta =3$, $m \approx 3$ eV and
$M_{\ast} = 20$ TeV. This gives $N^2 \approx 4000$ 
and $m _{\nu} ^2 \approx 2.2 \times 10^{-3}$ (eV)$^2$.

As seen in the figure,
there 
is a strong enhancement
of the decay width to LH $\tau$ 
for small $\tan \beta$ so that we get 
$O(1)$ BR to $\tau$, 
much larger than
in the $2$HDM-II. 
Also, the 
decay width to RH $\tau$ is strongly suppressed 
\footnote{
It is clear from Eq. (\ref{tauR}) that
the
decay width to RH $\tau$ is suppressed by the smaller of the two
factors $N^2$ and $\left( M_{\ast} / m_H \right) ^{\delta - 2}$,
which in this example is $\left( M_{\ast} / m_H \right) ^{\delta - 2}
\sim 100$.
}
so that the $\tau$ polarization is dominantly LH. 

If $\delta > 3$, then, for both cases (i) (a) and (b), 
the decay width to LH $\tau$ is small since
it is $\propto \left( m / M_{\ast} \right) ^2 \left( m_H / M_{\ast} \right)
^{\delta}$.
On the other hand,
in case (i) (a), the decay width to RH $\tau$ is suppressed
by $\sim N^2$ and thus this suppression remains about the same.
In case (i) (b), the decay width to RH $\tau$ is suppressed
by $\sim
\left( M_{\ast} / m_H \right)^ {\delta -2}$
and so the suppression becomes stronger.

A few comments on the case (i) (b) are in order.
It is clear that, for a given value of $M_{\ast}$ and $\delta$,
this effect occurs only for a specific value of the physical
neutrino mass, i.e,
$m_{\nu} \approx
m_{\nu}^{\hbox{\scriptsize max}}$.
Also, $m \sim$ (few) eV and
$M_{\ast} \sim 20$ TeV implies that the dimensionless ($5D$)
coupling, $\lambda$, is much larger than $1$ (see Eq. (\ref{m})). 
However, from the point of view of
the effective $4D$ theory, the Yukawa coupling
is $\lambda \; M_{\ast} / M_{Pl}$, which is much smaller than $1$ so
that the $4D$ theory is still perturbative. 
The $O(mR)^2$ correction to the normalization factor 
(Eq. (\ref{norm1})) is very large and so one might
worry about the higher order
corrections $\sim \sum _n (mR)^4
/n^4$, but these are suppressed by $\sim m^2 / M_{\ast}^2$ 
relative to the $O(mR)^2$ corrections and hence can be neglected.

For $M_{\ast} \sim 20$ TeV and $\delta \geq 3$, we get
$1/R \stackrel{>}{\sim} 10$ keV so that the SN$1987$a constraint is 
satisfied, although the $\delta =3$ case is marginal. 

{\bf Case (ii)} $M_{\ast} \gg 10$ TeV (decoupling limit)

In this case we have $\left( m_{\nu}^{\hbox{\scriptsize max}} \right) 
^2 \gg \Delta m ^2
_{\hbox{\small atm}}$ (roughly independent of $\delta$)
so that, to
get $m_{\nu} \sim 
\sqrt{ \Delta m ^2 _{\hbox{\small atm}} }$, we require $m \sim 
\sqrt{ \Delta m ^2 _{\hbox{\small atm}} }$.
As a result, the decay width to LH $\tau$ is negligible
since it is $\propto 
\left( m / M_{\ast} \right) ^2 \left( m_H / M_{\ast} \right)
^{\delta}$. Also,
$N ^2 \approx 1$ and so BR to RH $\tau$ is about the same as in 
$2$HDM-II. 
Thus we recover
the $2$HDM-II predictions in this case.

Of course,
in the case of degenerate neutrinos, $m_{\nu _{\mu}} \approx
m_{\nu _{\tau}}$, it is possible that $m_{\nu _{\tau}}^2 \gg
\Delta m_{\hbox{\small atm}}^2$, with
$|m_{\nu _{\tau}}^2 - m_{\nu _{\mu}}^2|
\sim \Delta m_{\hbox{\small atm}}^2$ to account for
the atmospheric neutrino anomaly.
For example, one can choose $M_{\ast} \sim 100$ TeV
and $m \; (\hbox{for} \;
\nu _{\tau}) \sim$ eV or larger.
This results in $N^2 \sim O(1-10)$ or $N^2 \gg 1$,
$m_{\nu _{\tau}} \sim m_{\nu}^{\hbox{ \scriptsize max}}
\sim $ eV and a phenomenology similar
to case (i) (a) or (b).
However, in this case, $m_{\nu _{\mu}}^2
\sim$ (eV)$^2$ also
which might be ruled out by the limit from BR$\left( \mu \rightarrow
e \gamma \right)$ \cite{fp} (depending
on $\nu _e - \nu_{\mu}$ mixing).

For $M_{\ast} \gg 10$ TeV and $\delta \geq 3$, we get
$1/R \gg 10$ keV so that the SN$1987$a constraint is easily
satisfied. 

{\bf Case (iii)} $M_{\ast} \sim$ (few) TeV.

In this case, $\left( m_{\nu}^{\hbox{\scriptsize max}}
\right)^2 \ll \Delta m ^2 _{\hbox{\small atm}}$
and the relevance to atmospheric neutrino oscillations
is not clear. Of course, it is possible that
$m_{\nu _{\tau}}^2 \ll m_{\nu _{\mu}}^2 \sim \Delta m^2 _{\hbox{\small atm}}$
(``inverted hierarchy''), but this scale for
$m_{\nu _{\mu}}$ has to be obtained by some other mechanism.

Nevertheless, as discussed earlier,
for $m ^2 \sim 10^{-2}$ (eV)$^2$, the decay width
to LH $\tau$ is very strongly enhanced so that the BR for $H^-$ decay to
{\em left-handed} $\tau$ can be $O(1)$ (even for moderate
$\tan \beta$ or
large $\delta$),
whereas the decay width to RH $\tau$ is 
suppressed compared to
$2$HDM-II 
so that the $\tau$
polarization is $\sim 100 \%$ LH. 
However, for $\delta =3$, we get
$1/R \sim 100$ eV and hence the SN$1987$a bound is violated. Whereas, for
$\delta \geq 4$, this constraint is satisfied.

\vskip .2in

To summarize,
we see that 
the following observations
at hadron/lepton colliders
can be a smoking gun signal for RH neutrino
in large extra dimensions: 

{\bf 1}. $\tau$'s from charged Higgs decays are dominantly 
{\em left-handed} as
in case (iii): $M_{\ast} \sim$ TeV,
$m^2 \sim 10^{-3}$ (eV)$^2$, case (i) (a)
$M_{\ast} \sim 10$ TeV, $m \sim 0.1$ eV, for small
$\tan \beta$, and case (i) (b):
$M_{\ast} \sim 10$ TeV, $m \sim$ eV

{\bf 2}.
suppressed (relative to
$2$HDM-II)
BR for charged Higgs decay to RH $\tau$ as in case (iii) and
case (i): $M_{\ast} \sim 10$ TeV
and $m^2 \stackrel{>}{\sim} 10^{-3}$ (eV)$^2$

{\bf 3}. BR$\left( H^- \rightarrow \tau \bar{\nu} \right)$ different
than that in $2$HDM-II. In particular, for small $\tan \beta$,
this BR is small
in $2$HDM-II whereas in cases (i) (b) and (iii), 
BR$\left( H^- \rightarrow \tau \bar{\nu} \right)$ can be $O(1)$.
 
Of course, the effect is sensitive to the values of $m_H$,
$\tan \beta$ and $\delta$.

\vskip .2in
If $m^2$ for 
$\nu _{e, \mu}$ are of the same order as in the case (i) (b) or
(iii),
then the
charged Higgs decays to LH $e$, $\mu$ can also be enhanced 
(depending as usual on the parameter values) -- of course,
charged Higgs decays to RH $e$, $\mu$ are negligible due to the small
Yukawa couplings.
In fact, the case
$M_{\ast} \sim$ (few) TeV can give
$m_{\nu _ {e, \mu}}$ of the correct order for
solar neutrino oscillations. However, 
these scenarios might be constrained 
(depending on the $\nu _e -
\nu _{\mu}$ mixing) by the limit on 
BR$\left( \mu \rightarrow e \gamma \right)$ \cite{fp}.

{\bf Singlet neutrino in sub-spaces}

It is also possible that the singlet neutrino lives in smaller number
of extra
dimensions, $\delta _{\nu} < \delta$, than the graviton \cite{aetal}.
For simplicity, assume that all the 
extra dimensions are of the same size, $R$. In this case, the effective
$4D$ neutrino Yukawa interaction is
\begin{equation}
S_{\hbox{int}} = \int d^4x \; \sum _n \; \frac{\lambda}{\sqrt{\left(
R M_{\ast} \right) ^{\delta _{\nu}} } }
l (x) h^{\ast} (x) \psi ^{(n)} (x).
\end{equation}
Using Eq. (\ref{funda})
we get
\begin{equation}
S_{\hbox{int}} = \int d^4x \; \sum _n \; \lambda
\left( \frac{M_{\ast}}{M_{Pl}} \right) ^ {\delta _{\nu} / \delta}
l (x) h^{\ast} (x) \psi ^{(n)} (x).
\end{equation}
Similarly, in the expressions for the charged Higgs decay widths to 
$\tau _L$ (Eqs. (\ref{tauL}) and (\ref{x}))
and $\tau _R$ (Eqs. (\ref{tauR}) and (\ref{norm1}))
and in the relation between $m$ and
the physical neutrino mass, $m_{\nu}$
(Eq. (\ref{mnu1})), we do the replacements:
\begin{eqnarray}
\delta & \rightarrow & \delta _{\nu} \nonumber \\
\left( \frac{M_{Pl}}{M_{\ast}}
\right) ^2 & \rightarrow & \left( 
\frac{M_{Pl}}{M_{\ast}} \right) ^{2 \left( 
\delta _{\nu} / \delta \right)}.
\end{eqnarray}

In this case, the maximum value of the physical neutrino mass
for given $M_{\ast}$, $\delta _{\nu}$ and $\delta$ is given by
\begin{equation}
m_{\nu}^{\hbox{\scriptsize max}} \approx M_{\ast} \left(
\frac{M_{\ast}}{M_{Pl}} \right) ^
{\delta _{\nu} / \delta}
\sqrt{ \frac{ \Gamma (\delta _{\nu} / 2) \; (\delta _{\nu} - 2) }
{ 2 \pi ^{ \delta _{\nu} / 2 } } },
\end{equation}
which is larger than in the case $\delta _{\nu} = \delta$ (Eq.
(\ref{mnumax})). For example,
for $M_{\ast} \approx 1$ TeV,
$\delta =6$ and $\delta _{\nu} =5$, we get
$m_{\nu}^{\hbox{\scriptsize max}} \sim
\sqrt{\Delta m^2 _{\hbox{\small atm}}}$. 
Thus, we can get $m_{\nu}^2 \sim \Delta m^2 _{\hbox{\small atm}}$
even for $M_{\ast}\sim $ a few TeV (unlike in the case 
earlier). For $m \sim 
\sqrt{ \Delta m^2 _{\hbox{\small atm}} }$,
we get $N^2 \sim O(1-10)$ and the charged Higgs decays to
$\tau$ are similar to the case (i) (a), whereas for
$m \sim$ eV, we get $N^2 \gg 1$ in which case
the charged Higgs phenomenology is similar to the case
(i) (b). For smaller values of
$\delta _{\nu} / \delta$ (for example, $\delta _{\nu} =3$, $\delta =4$)
and $M_{\ast} \sim$ TeV, we get 
$\left( m _{\nu}^{\hbox{\scriptsize max}} \right)^2 \stackrel{>}{\sim}$ 
(eV)$^2
\gg \Delta m ^2 _{\hbox{atm}}$. Thus, 
to get large decay width to
$\tau _L$, we require $m \stackrel{>}{\sim}$ eV
which results in $m_{\nu} ^2 \gg \Delta m^2 _{\hbox{\small atm}}$.

Next, we study the effects of virtual charged Higgs
in the scenario of RH neutrino in large extra dimensions.

\vskip .1in

{\large \bf Effects of virtual charged Higgs}

\vskip .1in

1. {\bf $B$ meson decays: $B^- \rightarrow l \bar{\nu}$}

\vskip .1in

The decay width in the SM is
\begin{equation}
\Gamma  _{W} \left( B^- \rightarrow l_R
\bar{\nu} \right) \approx
\frac{1}{8 \pi}
m_B m^2_l
\; f^2_B \; | V _{ub} | ^2 G^2_F \; \left( 1 -
\frac{m^2_l}{m_B^2} \right) ^2,
\label{BmuSM}
\end{equation}
where $i f_B p_{\mu} = \langle 0 | \bar{b} \gamma _{\mu} \gamma _5 u
| B^+ ( {\bf p} ) \rangle$ and $l = e$, $\mu$, $\tau$.

In $2$HDM-II with RH neutrino
in large extra dimensions, the charged Higgs exchange results
in a decay width
\begin{equation}
\Gamma _{H} \left( B^- \rightarrow l_L 
\psi \right) \sim \frac{1}{8 \pi}
m_B^3 \left( \frac{m \; m_b}{m^2_H} 
\right) ^2 \; f^2_B \; | V _{ub} |^2 G^2_F
\left(
\frac{m_B}{M_{\ast}} \right) ^{\delta} \left( \frac{M_{Pl}}{M_{\ast}}
\right)^2,
\label{BmuHiggs}
\end{equation}
where the last two factors are from the multiplicity of KK states
as usual.
There is also a charged Higgs exchange contribution
to the decay to RH $l$ (which is
significant only for $\tau$), but this is not enhanced by the multiplicity
factor and hence is not as important.

However, the larger effect of RH neutrino living
in extra dimensions
on the $B^- \rightarrow l \bar{\nu}$ decay width is due
to modification of the $W$ exchange amplitude as follows.
The SM neutrino
has a small mixture
$\sim m R /n$ of the heavier neutrinos, $\chi^{(n)}$'s, which
have Dirac masses
$\sim n / R$
(with the $\psi ^{(n)}$'s) 
(see Eq. (\ref{M})).
Due to this
effect, 
the virtual $W$ can decay into
LH $l$ and RH neutrino, $\psi ^{(n)}$'s. \footnote{
Or more directly, the chirality flip $m$ converts $\nu$ into
$\psi ^{(n)}$: see Eq. (\ref{M}).}
The same effect in $\pi^- \rightarrow e \bar{\nu}, \; \mu 
\bar{\nu}$ decays was considered
in \cite{dk}. 
The decay width $\Gamma _W \left( B^- \rightarrow l _L \psi \right)$
is (up to phase space factors) independent of the charged lepton
mass. The present experimental limit on BR$
\left( B^- \rightarrow \tau ^ -
\bar{\nu} \right)$ is weaker than that on BR$
\left( B^- \rightarrow
\mu ^ -
\bar{\nu} \right)$. In the case of $\nu _e$,
the ratio of $\pi^- \rightarrow e \bar{\nu}$
to $\pi^- \rightarrow \mu \bar{\nu}$ decay widths
\cite{dk} gives a much stronger constraint on $M_{\ast}$ than
BR $\left( B^- \rightarrow e_L
\psi \right)$. 
For $\delta =2$,
the effect of KK neutrinos
on the ratio of $\pi^- \rightarrow e\bar{\nu}$
to $\pi^- \rightarrow \mu \bar{\nu}$ decay widths
depends mostly on $m_{\nu _e}$
and hence does not constrain the
$m_{\nu _{\mu}}$ case as much.
So, we consider $B^- \rightarrow \mu _L \psi$ decay for which
\begin{eqnarray}
\Gamma _W \left( B^- \rightarrow \mu _L \psi \right)
& \approx & \frac{1}{8 \pi}
G_F^2 m_B f_B^2 | V_{ub} |^2 \sum_n \left( \frac{n}{R} \right) ^2 
\left( \frac{m R}{n} \right) ^2  
\frac{1}{N^2} \left( 1 - \frac{n^2 / R^2}{m_B^2}
\right) ^2 \nonumber \\
 & \approx & \frac{1}{8 \pi}
G_F^2 m_B f_B^2 | V_{ub} |^2 m_{\nu}^2 \left(
\frac{m_B}{M_{\ast}} \right) ^{\delta} \left( \frac{M_{Pl}}{M_{\ast}}
\right)^2 \; x_{\delta}, 
\label{Bmumix}
\end{eqnarray}
where $N^2$ is the normalization factor, Eq. (\ref{norm1}),
and we have used $m_{\nu} \approx m / N$ in the second line.
As usual, the factor $x_{\delta}$, Eq. (\ref{x}),
in the second line comes from converting the sum over KK states
to an
integral and the phase space integral, where we have neglected $m_{\mu}$.
Note that this effect depends on the $\mu$ neutrino mass and is
independent of $m_H$.
We can see that this
effect is larger than the charged Higgs effect (Eq. (\ref{BmuHiggs}))
by a factor $\sim \left( m_H / m_B \right)^4$.

The total decay width of the $B$ meson,
$\Gamma _B$, can be approximated by the decay width
for the $b$ quark decay, $b \rightarrow c \; (
\bar{c}s, \; \bar{u}d, \; l \bar{\nu})$:
\begin{equation}
\Gamma _B
\approx \frac{1}{192 \pi^3} |V_{cb}|^2 G_F^2 m_b ^5 \times 6,
\label{Btotal}
\end{equation}
where we have included a 
factor of $6$ for the number of virtual $W$ decay modes
weighted by the phase space.

We use $f_B / m_B \approx 1/20$ and $|V_{ub} / V_{cb}| \approx 0.1$.
Then,
using Eqs. (\ref{BmuSM}) and (\ref{Btotal}), we see that
BR($B^- \rightarrow \mu _R \bar{\nu})$ \footnote{This decay width is
actually reduced compared to the SM for the same reason as in the case
of $H^ - \rightarrow \tau _R \bar{\nu}$.}
is smaller than the present experimental
limit $\approx 2 \times 10^{-5}$
\cite{pdg}
by a factor of $O(100)$.
So, to obtain a constraint on $M_{\ast}$, we require that
decay width $B^- \rightarrow \mu_L \psi$ (Eq. (\ref{Bmumix}))
result in a BR 
smaller than the present limit. This gives
\begin{eqnarray}
\frac{ \Gamma _W \left( B^- \rightarrow \mu _L \psi \right)
}{\Gamma _B} 
& \approx & 
x \; \frac{24 \pi^2}{6} \left( \frac{f_B}{m_B}
\right)^2 \left( \frac{| V_{ub} |}{| V_{cb}|} \right) ^2 
\left( \frac{m_{\nu}}{m_B} \right) ^2 
\left( \frac{m_B}{M_{\ast}} \right) ^{\delta} \left( \frac{M_{Pl}}{M_{\ast}}
\right)^2 \nonumber \\
 & \stackrel{<}{\sim} & 2 \times 10^{-5}. 
\end{eqnarray}
Consider the case $\delta =2$.
Using Eq. (\ref{mnumax}), we see that for $M_{\ast} \approx 30$ TeV,
we get
$\left( m_{\nu}^{\hbox{\small max}} \right)^2 \approx 10^{-3}$ (eV)$^2
\sim \Delta m_{\hbox{\small atm}}^2$ so that to get 
$m_{\nu}^2 
\sim \Delta m_{\hbox{\small atm}}^2$,
we require $M_{\ast} \stackrel{>}{\sim} 30$ TeV.
\footnote{For this scale, $1/R \sim 0.5$ eV. Thus, if
$m^2 \sim 10^{-3} - 10^{-2}$ (eV)$^2$, then $mR$ is still 
smaller than $1$ and BBN constraint can also be evaded.
However, the SN$1987$a bound is violated.} 
Then, the above formula shows that BR $\left( B^- \rightarrow
\mu _L \psi \right)$ is below the present limit
by a factor of
$\approx 3$.
Thus, if the ongoing $B$-physics experiments improve the 
experimental limit by a factor of $\approx 3$, then the constraint
on $M_{\ast}$ from BR$\left(B^- \rightarrow \mu _L \psi \right)$
will become significant.
For $\delta \geq 3$, the minimum value of $M_{\ast}$ 
required to get
$m_{\nu}^2 \sim \Delta m_{\hbox{\small atm}}^2$ is a bit smaller, but 
BR $\left( B^- \rightarrow
\mu _L \psi \right)$ is suppressed by an extra factor of $ \left(
m_B / M_{\ast} \right) ^{\delta -2}$ so that the current
experimental bound
is easily
satisfied. Whereas for $\delta \geq 3$, 
the effect of KK neutrinos on
the ratio of $\pi^- \rightarrow e \bar{\nu}$
to $\pi^- \rightarrow \mu \bar{\nu}$ decay widths depends
on $\Delta m^2 _{e - \mu}$ and gives a strong constraint
on $M_{\ast}$ \cite{dk}. 

\vskip .1in

2. {\bf Lepton flavor violating decays: $\mu \rightarrow e \gamma$}

\vskip .1in

There is also 
a charged Higgs exchange contribution at one-loop
to flavor violating decays,
for example, $\mu \rightarrow e \gamma$. 
The one-loop
$W$ exchange contribution to this decay with
RH neutrino in large extra dimensions
was studied in \cite{fp}.
As usual, this decay is enhanced by the large number of RH
neutrino KK states in the loop -- this multiplicity factor
compensates for the small neutrino mass (or Yukawa coupling). 
For $\Delta m^2 _{e - \mu} \sim 10^{-5}$ (eV)$^2$ and
$\theta _{e - \mu} \sim \pi /4$, the limits on $M_{\ast}$ are
$\sim 100$ TeV and $\sim 35$ TeV for 
$\delta = 2$ and $3$, respectively.  

The dominant
contribution (with a single Higgs doublet) 
comes from longitudinal $W$ boson 
and heavy KK states (with
masses close to the UV cut-off, $M_{\ast}$)
in the loop 
\cite{fp} (at least for $\delta \geq 3$). Therefore,
in  $2$HDM-II, 
the contribution with charged Higgs in the loop 
is
of the same order as
the $W$ boson exchange effect 
as long as
$m_H <$ the cut-off, $M_{\ast}$. 

\vskip .1in

3. {\bf Top quark decay: $t \rightarrow b \tau^+ \nu$}

\vskip .1in

The SM decay $t \rightarrow b W^+$ has width
\begin{eqnarray}
\Gamma \left( t \rightarrow b W^+ \right) & \approx &
\frac{G_F m^3_t}{8 \pi \sqrt{2}}
|V_{tb}|^2 (1 - x_W^2) (1 + x_W ^2 - 2 x_W ^4) \nonumber \\
 & \approx & 0.6 \frac{G_F m^3_t}{8 \pi}
\end{eqnarray}
where $x _W \equiv m_W / m_t \approx 1/2$ and $|V_{tb}| \approx 1$. 
The decay width to $b \tau^+ \nu$ is smaller
by a factor of $1/9$.
 
The (anti-)top quark 
decay to LH $\tau$ with charged Higgs exchange is enhanced
by the large number of RH $\nu$ KK states. For small $\tan \beta$, we get:
\begin{equation}
\Gamma \left( \bar{t} \rightarrow \bar{b} 
\tau _L \psi \right) \sim \frac{\sqrt{2} \; G_F m^3_t}{
192 \pi^3} \left( \frac{m}{v} \right)^2 \cot ^4 \beta
\left( \frac{m_t}{m_H} \right) ^4
\left( \frac{m_t}{M_{\ast}} \right) ^{\delta}
\left( \frac{M_{Pl}}{M_{\ast}}
\right)^2.
\end{equation}

Thus, the ratio of the top decay widths to $b
\tau ^+ \nu$ due to charged Higgs exchange and that in the SM
is
\begin{equation}
\sim \frac{9 
\; \sqrt{2}}{0.6 \; 24 \pi^2 } \left( \frac{m}{v} \right)^2 \cot ^4 \beta
\left( \frac{m_t}{m_H} \right) ^4
\left( \frac{m_t}{M_{\ast}} \right) ^{\delta}
\left( \frac{M_{Pl}}{M_{\ast}}
\right)^2
\end{equation}
which
is $O(1)$ 
for 
$m^2 \sim 10^{-2}$ (eV)$^2$,
$\tan \beta \sim 2$, $M_{\ast} \sim 2$ TeV, $\delta = 3$ and $m_H 
\sim 200$ 
GeV, i.e., the top decay width
to $b \tau ^+ \nu$ can be enhanced by $O(100 \%)$.
\footnote{Of course, for $M_{\ast} \sim 2$ TeV, the physical neutrino mass
is smaller than $\sqrt{ \Delta m^2 _{\hbox{\small atm}}}$ and also
the SN$1987$a bound is violated for $\delta = 3$.} 
At the LHC, it should be possible to measure the top BR's at the
$\sim 10 \%$ level so that the above effect can be observed. 
If the three SM neutrinos masses are (roughly) of the
same order, then the same
effect will be observed in top quark decays to $b \mu^+ \nu$ and $b e^+
\nu$.
On the other hand, if $m_{\nu_{\mu, e}} \ll m_{\nu _{\tau}}$, then the top
quark decays to $e^+$, $\mu^+$ will be unaffected --
this violation of lepton universality in top quark decays
can be observed at the LHC or even in
Run II of the Tevatron.

In summary, we have studied the effects of RH neutrino living in 
large extra dimensions on charged Higgs phenomenology. We have shown
that  
in this model charged Higgs decays to {\em left-handed} $\tau$
can be enhanced, with $O(1)$ branching ratio, and
decays to RH $\tau$ can be
suppressed. 
Constraints from SN$1987$a energy loss indicate that effects of bulk neutrino
cannot be observed in atmospheric
neutrino oscillations \cite{bcs}; 
however, we have seen that the effects in charged
Higgs decays can still be observed.
Thus, charged Higgs decays can provide signatures
for these models.

\end{document}